# MoS$_2$ Heterostructure with Tunable Phase Stability: Strain Induced Interlayer Covalent Bond Formation

Bin Ouyang[1,*], Shiyun Xiong[2], Zhi Yang[3,4], Yuhang Jing[5,6], Yongjie Wang[7]

1. National Center for Supercomputing Applications, University of Illinois at Urbana–Champaign, Urbana, Illinois 61801, United States.
2. Functional Nano and Soft Materials Laboratory (FUNSOM) and Collaborative Innovation Center of Suzhou Nano Science and Technology, Soochow University, Suzhou, JiangSu 215123, China.
3. College of Physics and Optoelectronics, Taiyuan University of Technology, Taiyuan, ShanXi 030024, China.
4. Key Lab of Advanced Transducers and Intelligent Control System, Ministry of Education and Shanxi Province, Taiyuan University of Technology, Taiyuan, ShanXi 030024, China.
5. Department of Astronautical Science and Mechanics, Harbin Institute of Technology, Harbin, Heilongjiang 150001, China.
6. Beckman Institute for Advanced Science and Technology, University of Illinois at Urbana-Champaign, Urbana, Illinois 61801, United States.
7. Department of Electrical Engineering and Computer Engineering, University of Michigan, Ann Arbor, Michigan 48109, United States.

**Abstract**

Due to the distinguished properties offered by different structural phases of monolayer MoS$_2$, phase engineering design are urgently required for achieving switchable structural phase. Strain engineering is widely accepted as a clean and flexible method, however, can not be achieved in engineering monolayer MoS$_2$ phase transition because the critical biaxial strain required (~15%) is much larger than measured elastic limit (~11%). In this study, employing density functional theoretical calculations, it has been found out that with the forming of heterostructure between MoS$_2$ with buckled 2D materials such as silicence, germanene and stanene, only a small strain can trigger the phase transition. As being suggested by the constructed phase stability diagram, biaxial deformation as low as 3% in MoS$_2$/silicene and MoS$_2$/stanene sandwich structure, would be sufficient to induce the structural phase transition in MoS$_2$ lattice. This strain falls well within experimental elastic limit, thus would be feasible to realize in experiment. The origin of such behavior can be understood as strain induced interlayer covalent bond formation, which finally make MoS$_2$ lattice more sensitive to external strain. This theoretical work provides one realistic route for achieving flexible phase stabilities in experimental design.

* Author to whom correspondence should be addressed. E-Mail: bouyang@illinois.edu

# 1. Introduction

In addition to the promising properties offered by $MoS_2$ in the field of optoelectronics[1, 4, 29, 30], piezoelectronics[33, 39] and valleytronics[18, 19], the probing of polytypic structures has recently inspired several new possibilities[11, 15, 21-23]. With various experimental designs[11, 13, 15], structural phase transition will occur on the original 2H phase accompanied with a shift of S sub lattice, which forms the 1T phase. However, due to the instability of 1T lattice without support, peierls distortion would always take place and further leads to the 1T→ 1T' (1T'') phase transitions[21, 26]. The polymorphism of $MoS_2$ will enrich its electronic properties and lead to new applications. As can be inferred from the band structure, $1T-MoS_2$ is metallic and has been used for hydrogen evolution[17, 34, 37] and lithium/sodium batteries[20, 35, 37], while $1T'-MoS_2$ and $1T''-MoS_2$ possess narrow band gaps ( < 0.1 eV) with flexible tunability under external field, which is promising for low dimensional topological insulator design[5, 26, 36].

Up to date, means of achieving flexible phase transitions in $MoS_2$ and its nanostructures are largely limited to charge based methods[11, 15, 21, 23, 32], which introduce various defects into the membranes in the meantime[11, 15]. The ferroelastic behavior as being studied by Duerloo et al[7], Li et al[14] and Ouyang et al[21], inspires the potential of utilizing strain engineering as alternative option which are relatively cleaner and more flexible. However, according to first principle calculations[7, 21], the critical elastic planar strain required for triggering 2H→T transition in $MoS_2$ monolayer turns out to be 0.14 for biaxial deformation and >0.2 for uniaxial deformation[7, 21], those are significant larger than its elastic limit reported experimentally[2]. As a result, with the strain loaded on MoS$_2$, the membrane will be broken before phase transition can happen. That greatly inhibit the realization of strain engineering for flexible phase transitions.

Recently, building 2D heterostructure has attracted tremendous attention as an effective way to tailor properties of 2D materials[12, 22-24, 28, 33]. This gives a hint for engineering phase stability of MoS$_2$ through heterostructure design. In this paper, we are interested in combining MoS$_2$ with 2D materials consisting buckled hexagonal lattice since it will ensure more orbital overlapping at interface. As a result, silicene[31], germanene[6] and stanene[38] are selected and it appears that the structural phase transition of MoS$_2$ can be flexibly tuned by strain well below its elastic limit. With the analysis of charge transfer, the mechanism has been discovered to be originated from strain induced covalent interaction between MoS$_2$ with the 2D bucked materials. Unlike the graphene derivatives with only planar orbitals[22, 28], the covalent bridge can be directly formed between MoS$_2$ and 2D bucked monolayers (BML) due to orbital overlapping. Accompanied by the forming of covalent bonding, the phase stability of T phases will be enhanced and therefore 2H→T structural transition will occur with much smaller planar strain than that in the free standing MoS$_2$. On base of the theoretical predictions, a new route for tuning phase stability of MoS$_2$ nanostructure combining interface and lattice strain has been suggested. Meanwhile, due to the distinguished properties between 2H-MoS$_2$ or T-MoS$_2$ integrated with BMLs, this study will also shed lights on designing 2D heterostructures switchable properties.

## 2. Methodology

### 2.1. Modeling of Atomic Structures.

Due to the lattice mismatch in MoS$_2$ with BMLs lattices, supercells consist of different units of MoS$_2$ and BMLs are selected to minimize the lattice distortion. To minimize misfit strain, 6×6, 10×10, 6×6 supercell of MoS$_2$ are selected for constructing heterostructures with silicene, germanene and stanene correspondently, while the supercell size of BMLs is set as 5×5, 8×8 and 4×4 separately for silicene, germanene and stanene (as illustrated by Figs. 1(a)-(c)). Since we are focusing on the structural transition of MoS$_2$ lattice, the 2H-MoS$_2$ lattice is set as a reference. With this structural model, the corresponding lattice mismatches are smaller than 1.3%, 1.0% and 2.0%. In this work, we have considered MoS$_2$/BML bilayer and BML/MoS$_2$/BML sandwich heterostructures, which account for interface interaction from one side and both side of MoS$_2$. Both types of heterostructures can be easily synthesized experimentally with deposition based methods[16, 27].

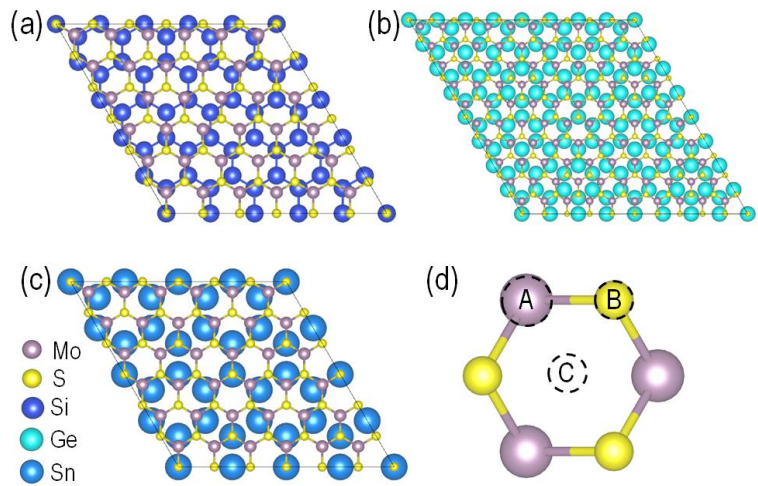

**Fig. 1 (Color Online)** Atomic illustration of supercell size adopted for (a) Si/MoS$_2$; (b)

Ge/MoS$_2$ and (c) Sn/MoS$_2$. Three types of sublattice site are showed in (d).

It is worth noting that due to the variation in orientation, there are different possibilities to stack the heterostructures. All those possibilities can be identified with the help of Fig. 1(d). As being showed by Fig. 1(d), three symmetric sites labelled as A, B and C exist in the hexagonal basal plane of MoS$_2$. In order to ensure the most orbital overlapping with monolayer MoS$_2$, the bonds among BML atoms (Si, Ge, Sn) are set perpendicular to either AB, AC or BC direction. As a result, there will be 6 stacking configurations for bilayer MoS$_2$/BML heterostructures and 21 stacking configurations for BML/MoS$_2$/BML sandwiched heterostructures. All the potential stacking configurations are listed in detail at Fig. S1 (Supporting Information).

**2.2. Structure optimization.**

First-principles DFT calculations employing the Perdew-Burke-Ernzerhof (PBE) functional[25] and projector augmented-wave (PAW)[3] method were performed using the Vienna ab initio simulation package (VASP) for structure optimization and energetic calculations. The dispersive van der Waals interactions between the MoS$_2$ and BMLs were included using the DFT-D2 method of Grimme[8-10]. In each calculation, up to 5×5×1 *k*-point grid and an energy cutoff of 400 eV were adopted. Both the k-grid and cutoff energy have been tested to be converged in energy. A vacuum space of 20 Å is used for each slab model to eliminate image interactions. When performing the structure optimizations, the system is regarded as converged when the force per atom is less than 0.01 eV/Å.

## 3. Results and discussion

Obtained from the first principle DFT calculations, the interface configurations for three types of bilayer MoS$_2$/BML heterostructures are demonstrated in Fig. 2. The interface interactions are illustrated by the deformation charge density shown in Figs. 2(a)-(c). In each panel, the figures from up to down correspond to the side view of the 2H-MoS$_2$/BML and the T-MoS$_2$/BML bilayer with the lowest energetic states. Since there is significant lattice distortion in T-MoS$_2$/BML heterostructures, the planar view of each monolayer at each T-MoS$_2$/BML heterostructure has also been included below the side view for more structural information.

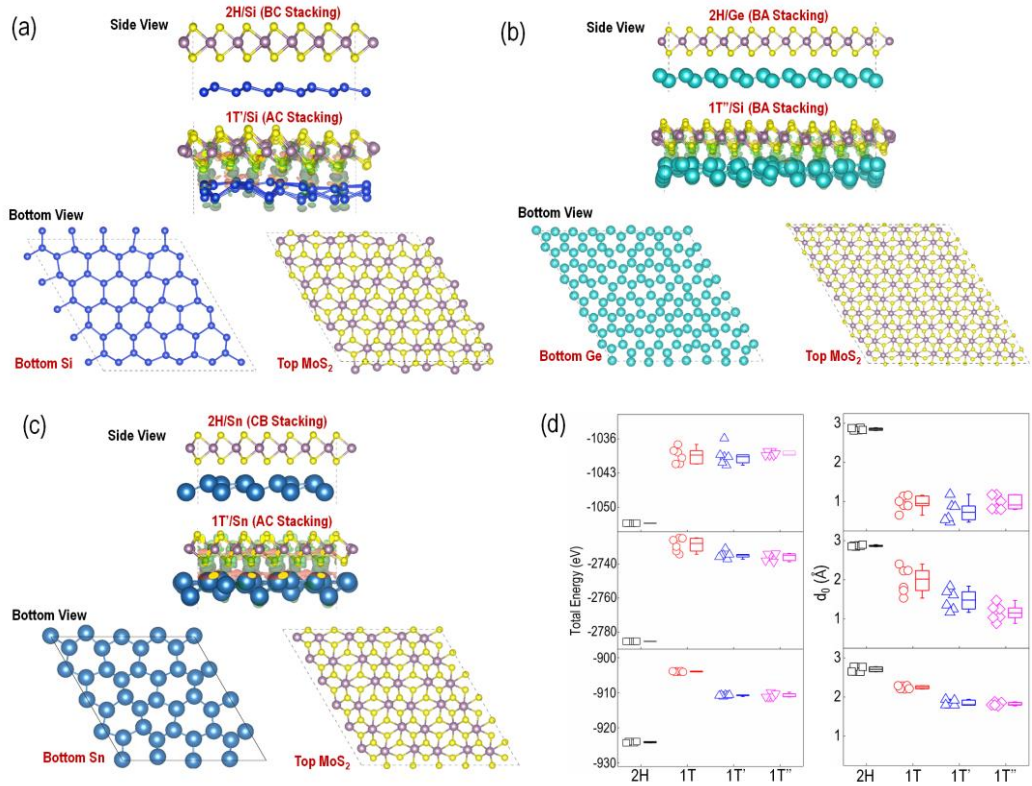

**Fig. 2 (Color Online)** Ground state and deformation charge density for bilayer structure of MoS$_2$ with (a) Si, (b) Ge, (c) Sn; In each figure from (a)-(c), the top two plots are side views of heterostructures with deformation charge density illustrated with the same isosurface threshold value. While the two subplots underneath are bottom view of two monolayers individually. (d) Box chart plot for the total energy $E_0$ and interlayer distance $d_0$ of different configurations; top: Si/MoS$_2$, middle: Ge/MoS$_2$ and bottom: Sn/MoS$_2$.

It can be inferred that for all three groups of MoS$_2$/BML heterostructures, the 2H-MoS$_2$ tends to interact with BMLs via van der Waals bonds since no charge transfer can be observed. On the contrary, the T-MoS$_2$ (1T, 1T', 1T'') tends to form covalent bonds with BMLs as evidenced by the significant charge transfer at the interface. The formation of covalent bonds between T-MoS$_2$ and BML can also be evidenced by significant lattice distortions perpendicular to the basal plane. Moreover, as shown in Figs. 2(a)-(c), compared to the buckling direction, the atomic position in the planar direction is almost unchanged and the lattice projection is kept as the hexagon shape.

Those phenomena observed in Figs. 2(a)-(c) can be further confirmed by the quantitative calculations of the total energy $E_0$ and interlayer distance $d_0$ shown in Fig. 2(d). For each system, the total energy as well as layer distance between MoS$_2$ and BML of all stacking possibilities are presented. The exact value for all stacking sequences can be found in the supporting information. For 2H-MoS2/BML heterostructures, $E_0$ and $d_0$ do not change too much within different orientations between MoS2 and BML layers. While these two quantities strongly depend on the relative layer orientations in T-MoS2/BML (T = 1T, 1T' and 1T'') heterostructures. The strong orientation dependency of energetics and interfacial distance indicates that there will be strong covalent interaction between T-MoS$_2$ and BML layers. Additionally, by comparing the interfacial distance $d_0$ between BML and different phases of MoS$_2$, it can be found that for all the three groups, 2H-MoS$_2$/BML always possess the largest interlayer distance (~ 3 Å). This also deliver the information that 2H-MoS$_2$/BML possess the weakest interlayer interaction.

**Tab. 1:** Calculated relative energy per unit cell $\Delta E$ for free standing MoS$_2$ and bilayer heterostructure, $\Delta E_{min}$ and $\Delta E_{avg}$ indicate the minimum and average $\Delta E$ among all stacking sequences.

| Structure | MoS$_2$ | MoS$_2$/Si | | MoS$_2$/Ge | | MoS$_2$/Sn | |
|---|---|---|---|---|---|---|---|
| Energy(eV) | $\Delta E$ | $\Delta E_{min}$ | $\Delta E_{avg}$ | $\Delta E_{min}$ | $\Delta E_{avg}$ | $\Delta E_{min}$ | $\Delta E_{avg}$ |
| 1T | 0.839 | 0.337 | 0.385 | 0.513 | 0.568 | 0.552 | 0.561 |
| 1T' | 0.551 | 0.330 | 0.383 | 0.483 | 0.509 | 0.370 | 0.371 |
| 1T'' | 0.634 | 0.385 | 0.396 | 0.468 | 0.498 | 0.356 | 0.371 |

In order to quantitatively determine the influence of interfacial bonds on phase stability of MoS$_2$, we have calculated the relative energy difference $\Delta E$ between the 2H and T phases per unit cell, which is defined as:

$$\Delta E = (E_T^\varphi - E_{2H}^\varphi)/N_{Unit} \tag{1}$$

in which $E_T^\varphi$ and $E_{2H}^\varphi$ refers to calculated total energies for the T phase (1T, 1T' or 1T'') and 2H phase, respectively. $N_{Unit}$ represents the number of MoS$_2$ unit in the superlattice. $\varphi$ can be either the bilayer heterostructure, the sandwich heterostructure or the free standing MoS$_2$. As shown in Tab. 1, compared to the free standing structures, $\Delta E$ will be decreased by forming a bilayer heterostructures with all the three considered BML, among which the change takes the order MoS$_2$/Ge < MoS$_2$/Sn < MoS$_2$/Si. The reduced $\Delta E$ indicates that the transition from 2H to T structures will be much easier when a bilayer heterostructure is formed. Consequently, this strategy can be cooperated with the strain engineering to tune among different phases. Since the energy difference is largely reduced, the required strain could be much smaller than that in the free standing MoS$_2$, thus opens a potential way to realize the phase transformation through deformation in experiments.

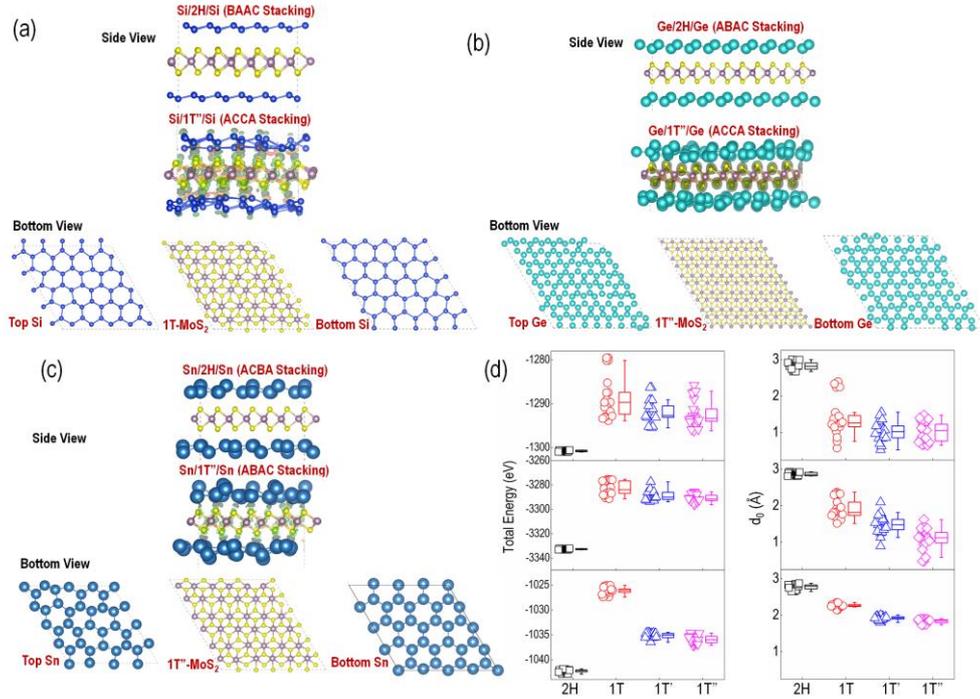

**Fig. 3 (Color Online)** Ground state and deformation charge density for sandwich structure of MoS$_2$ with (a) Si, (b) Ge, (c) Sn; (d) Calculated total energy and inter-layer distance for all the potential stacking of BML/2H/BML heterostructures. In each figure from (a)-(c), the top two plots are side views of the heterostructures with deformation charge density illustrated with the same isosurface value. While the three subplots underneath are the bottom view of the three monolayers individually.

The configurations of BML/MoS$_2$/BML sandwiched heterostructures are also examined and demonstrated in Fig. 3. Similar to the bilayer systems, BML interacts with T-MoS$_2$ from both sides via covalent interactions while interacts with 2H-MoS$_2$ via weak Van der Walls forces. Also, E$_0$ and d$_0$ of BML/T-MoS$_2$/BML structures highly depend on the stacking orientations while the fluctuation is even larger than the bilayer systems. In order to quantify the influence of interfaces on phase stability, the relative energy $\Delta E$ is also evaluated for each configuration and demonstrated in Tab. 2.

**Tab. 2:** Calculated relative energy per unit cell $\Delta E$ for the free standing $MoS_2$ and sandwiched heterostructure.

| Structure | $MoS_2$ | Si/$MoS_2$/Si | | Ge/$MoS_2$/Ge | | Sn/$MoS_2$/Sn | |
|---|---|---|---|---|---|---|---|
| Energy(eV) | $\Delta E$ | $\Delta E_{min}$ | $\Delta E_{avg}$ | $\Delta E_{min}$ | $\Delta E_{avg}$ | $\Delta E_{min}$ | $\Delta E_{avg}$ |
| 1T | 0.839 | 0.187 | 0.341 | 0.409 | 0.496 | 0.406 | 0.449 |
| 1T' | 0.551 | 0.153 | 0.248 | 0.399 | 0.440 | 0.167 | 0.198 |
| 1T'' | 0.634 | 0.131 | 0.218 | 0.362 | 0.418 | 0.138 | 0.178 |

As illustrated in Tab. 2, the relative energy difference $\Delta E$ is significant decreased in each of the sandwiched structure considered compared to the free standing $MoS_2$. This reduction is even more significant than that in the corresponding bilayer structures, indicating that the transition from 2H to T phases is even easier in sandwiched structure compared to the bilayer systems.

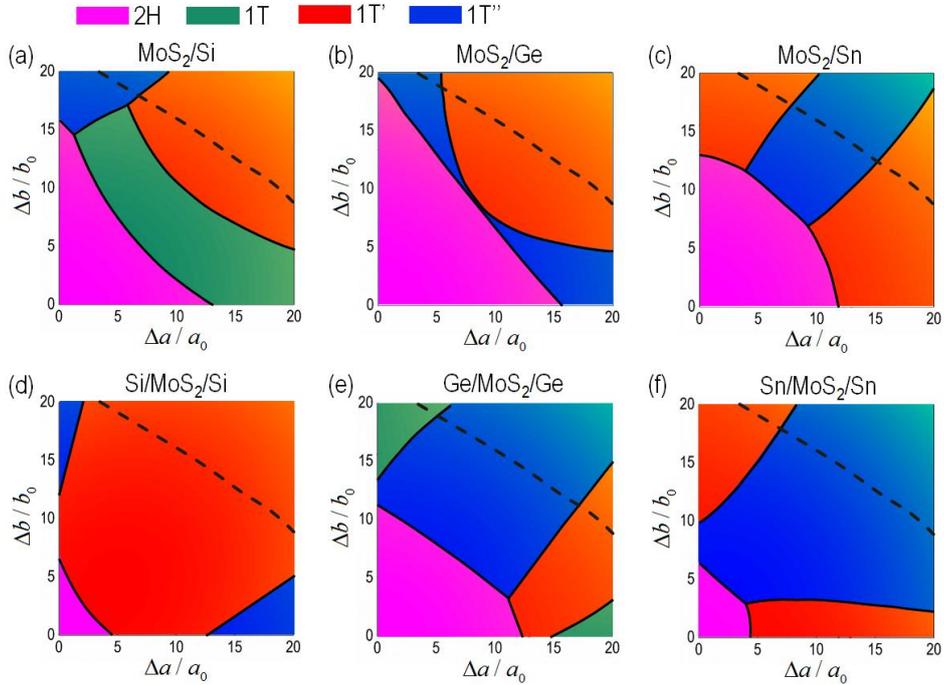

**Fig. 4 (Color Online):** Phase stability diagram for (a) $MoS_2$/Si; (b) $MoS_2$/Ge; (c) $MoS_2$/Sn; (d) Si/$MoS_2$/Si; (e) Ge/$MoS_2$/Ge; (f) Sn/$MoS_2$/Sn. The solid line indicates the phase boundaries predicted while the dashed line indicates the 2H/1T' phase boundary of freestanding monolayer $MoS_2$[7].

With the reduction of energy gap between 2H and T phases by building heterostructures, the phase transition will easily occur with the assistance of other environment variables. More specifically, it should require much smaller elastic strain than free standing $MoS_2$ for inducing the corresponding phase transitions. In order to provide a direct evidence for this idea, the phase stability diagram has been constructed for different $MoS_2$ heterostructures treating planar strain as state variables (Fig. 4).

As indicated in Fig. 4, the critical strain required for inducing 2H→T (1T, 1T' and 1T'') phase transition has been greatly reduced with the assistance of interface formed by $MoS_2$ and BMLs. Moreover, being consistent with the interaction analysis in Fig. 2-3 and Tab. 2-3, heterostructures formed by Germanene and $MoS_2$ need the highest critical strain to take the transition from 2H to T phases due to the weakest interlayer interactions. In the contrary, the critical strain of the heterostructures with silicene is the smallest thanks to the strong interlayer interactions. Meanwhile, with the enhanced interface overlapping by forming the sandwiched structures, the phase stability will be more sensitive to elastic strain compared to the bilayer structure. To facilitate the detailed comparisons, we have listed the critical lattice deformation required for 2H→T transition in Tab. 3. From this dataset, it can be clearly referred that the elastic strain required for 2H→T transitions in $MoS_2$ are almost halved with the forming of bilayer heterostructures. Furthermore, the designing of sandwich structure will bring the overall picture more appealing since only around 0.05 of uniaxial deformation and around 0.03 of biaxial deformation are required. As being suggested by experimental work, these range of strains would be easily achieved in a real-world design.

Tab. 3: Critical lattice deformation required for inducing 2H→T (1T, 1T', 1T") phase transition within $MoS_2$ lattice.

| Strain | $MoS_2$ | $MoS_2$/Si | $MoS_2$/Ge | $MoS_2$/Sn | Si/$MoS_2$/Si | Ge/$MoS_2$/Ge | Sn/$MoS_2$/Sn |
|---|---|---|---|---|---|---|---|
| $\Delta a / a_0$ | >0.2 | 0.13 | 0.15 | 0.12 | 0.05 | 0.12 | 0.05 |
| $\Delta b / b_0$ | >0.2 | 0.15 | 0.19 | 0.13 | 0.06 | 0.11 | 0.06 |
| Biaxial | 0.15 | 0.06 | 0.08 | 0.08 | 0.03 | 0.06 | 0.03 |

## 4. Summary

To conclude, flexible phase transition can be achieved in 2D $MoS_2$ by forming heterostructures with silicone, germanene and stanene. Due to the covalent bond formation between $MoS_2$ and BML, the charge transfer will diminish the energy gap between different phases of $MoS_2$. The reduced energy gap greatly decreases the critical strain that required for phase transition from the 2H to T phases. The critical biaxial strain can be as low as 3%, which is almost 5 times smaller than that in the free standing $MoS_2$. This small critical strain is well below the elastic limit, thus can be realized experimentally. Therefore, we propose that by designing heterostructures between $MoS_2$ and buckled monolayers, potential devices with strain switchable electronic properties can be developed. With the information provided by this study, potential new devices based on transition metal dichagonides 2D heterostructures can be developed.

## 5. Acknowledgement

The authors would like to thank the supports from National Natural Science Foundation of China (U1510132). Various computing resources including National Center for Supercomputing Applications in University of Illinois at Urbana–Champaign, Supercomputer Consortium Laval UQAM McGill and Eastern Quebec and Professor Xiong's group cluster in Soochow University are also greatly appreciated.